\documentclass[twocolumn,showpacs,preprintnumbers,superscriptaddress,amsmath,amssymb,floatfix,prl]{revtex4}

\usepackage{graphicx}
\usepackage{dcolumn}
\usepackage{bm}
\usepackage{soul}

\usepackage[utf8]{inputenc}
\usepackage{xcolor}

\newcommand{\lD}{\lambda_{\textrm{D}e}}
\newcommand{\vkO}{\vec{k}_0}

\newcommand{\kLW}{k_{\textrm{LW}}}
\newcommand{\vkLW}{\vec{k}_{\textrm{LW}}}
\newcommand{\vkLWp}{\vec{k}_{\textrm{LW}}^\prime}
\newcommand{\kEM}{k_{\textrm{EM}}}
\newcommand{\vkEM}{\vec{k}_{\textrm{EM}}}
\newcommand{\wSRS}{\omega_{\textrm{SRS}}}

\newcommand{\vkSRS}{\vec{k}_{\textrm{SRS}}}

\begin{document}


\title{A laser-plasma interaction experiment for solar burst studies}

\author{J.-R. Marquès}
\affiliation{
LULI, CNRS, CEA, Sorbonne Université, École Polytechnique, Institut Polytechnique de Paris, Palaiseau, France
}

\author{C. Briand}
\affiliation{LESIA, Observatoire de Paris, Université PSL, CNRS, Sorbonne Université, Université de Paris, Meudon, France
}

\author{F. Amiranoff}
\affiliation{
Sorbonne Université, LULI, CNRS, CEA, École Polytechnique, Institut Polytechnique de Paris, Paris, France
}
\author{S. Depierreux}
\affiliation{
CEA, DAM, DIF, F-91297 Arpajon, France}
\author{M. Grech}
\affiliation{
LULI, CNRS, CEA, Sorbonne Université, École Polytechnique, Institut Polytechnique de Paris, Palaiseau, France
}
\author{L.~Lancia}
\affiliation{
LULI, CNRS, CEA, Sorbonne Université, École Polytechnique, Institut Polytechnique de Paris, Palaiseau, France
}
\author{F. Pérez}
\affiliation{
LULI, CNRS, CEA, Sorbonne Université, École Polytechnique, Institut Polytechnique de Paris, Palaiseau, France
}

\author{A. Sgattoni}
\affiliation{LESIA, Observatoire de Paris, Université PSL, CNRS, Sorbonne Université, Université de Paris, Meudon, France
}
\affiliation{
Sorbonne Université, LULI, CNRS, CEA, École Polytechnique, Institut Polytechnique de Paris, Paris, France
}

\author{T. Vinci}
\affiliation{
LULI, CNRS, CEA, Sorbonne Université, École Polytechnique, Institut Polytechnique de Paris, Palaiseau, France
}

\author{C. Riconda}
\affiliation{
Sorbonne Université, LULI, CNRS, CEA, École Polytechnique, Institut Polytechnique de Paris, Paris, France
}

\date{\today}

\begin{abstract}
A new experimental platform based on laser-plasma interaction is proposed to explore the fundamental processes of wave coupling at the origin of interplanetary radio emissions. It is applied to the study of electromagnetic (EM) emission at twice the plasma frequency ($2\omega_p$) observed during solar bursts and thought to result from the coalescence of two Langmuir waves (LWs). In the interplanetary medium, the first LW is excited by electron beams, while the second is generated by electrostatic decay of Langmuir waves. In the present experiment, instead of an electron beam, an energetic laser propagating through a plasma excites the primary LW, with characteristics close to those at near-Earth orbit. The EM radiation at $2\omega_p$ is observed at different angles. Its intensity, spectral evolution and polarization confirm the LW-coalescence scenario.

\end{abstract}


\maketitle

Solar flares generate intense electromagnetic (EM) radiations in the radio domain (1-100 MHz) that are the signature of electron beams propagating in the interplanetary medium \citep{Pick08a,Reid14a}. Detected by space and ground-based radio telescopes, these EM waves could, in principle, provide characteristics of the electron beams, thus opening the prospect for direct applications in space weather. The individual steps resulting in such emission have been proposed in the 50’s \cite{Ginzburg1959,Willes1996}: the fast electron beams generated during solar flares provide the free energy necessary to destabilize the interplanetary plasmas leading, in particular, to the excitation of electron plasma waves (Langmuir waves, LW) through beam-plasma instabilities. These can produce EM waves at the local plasma frequency ($\omega_p$) or its harmonics. This paper focuses on the type-III radio bursts emitted at $2\omega_p$, which is thought to come from a two-step mechanism:

\begin{eqnarray}\label{eq:coupling_eq}
\textrm{LW}\longrightarrow \textrm{LW}^\prime + \textrm{IAW}^\prime\\
\textrm{LW}+\textrm{LW}^\prime \longrightarrow \textrm{EM}_{2\omega_p}\label{eq:coupling_eq2}
\end{eqnarray}
In step (\ref{eq:coupling_eq}), known as the Langmuir Decay Instability (LDI), 
the primary LW decays into a secondary one (LW$^\prime$, almost counter-propagating) and an ion acoustic wave (IAW$^\prime$). In step (\ref{eq:coupling_eq2}), known as Langmuir wave coalescence, the nonlinear coupling between the two LWs generates a current at $2\omega_p$, a source term for an EM wave at the same frequency \cite{Willes1996,Gorbunov2004}.

Numerous observations from space instruments have provided in-situ measurements of these mechanisms in the very specific plasma environment of the interplanetary medium. Recently, relations between the frequencies, wavevectors and phases of the waves involved in the wave coupling mechanism have been confirmed \citep{Henri09a,Briand14a}. However, these measurements remain limited spatially (single-point measurements of LWs) and temporally (due to the reduced telemetry). Analytical models and numerical simulations have been developed to interpret these space observations. Recently, three-dimensional EM particle-in-cell simulations tackled several questions regarding the efficiency of the conversion process from LW to EM waves, and the emission pattern of the EM waves  \citep{Thurgood2016,Henri19a}. These models and simulations yet remain to be confronted with experimental data.

Laboratory experiments provide a complementary and powerful method to study the fundamental mechanisms of wave coupling.
In the 1980's \citep{Whelan1981a,Michel1982a,Cheung1982a,Schneider1982a,Whelan1985} a few experiments based on the injection of an electron beam in a plasma, magnetized or not, have reported the detection of EM emission at $\omega_p$ or $2\omega_p$. The radiated powers at $\omega_p$ and $2\omega_p$ were respectively $\sim 10^{-6}$ and $\sim 10^{-9}$ of the total beam power. These pioneering laboratory experiments were all performed with low-energy electron beams ($< 3$ keV) in plasmas of low electron temperature ($T_e <$ 3 eV). As a result they all lead to very similar wave parameters. Only one of these experiments \citep{Whelan1981a} measured the polarization of the EM radiation, and none measured the angular pattern of the emission. Laboratory data in a larger range of plasma parameters, with a larger set of probed regions and a broader type of diagnostics would allow better comparisons with space data and to better exploit the new capabilities of 3D simulations.

Laser-generated plasmas provide today a unique and effective platform to study the generation of the 2$\omega_p$ emission. Indeed, in laser-plasma interaction the laser takes the role of the electron beam in the interplanetary medium: the primary LW can be generated by laser-stimulated backward Raman instability. With appropriate conditions, steps (\ref{eq:coupling_eq}) and (\ref{eq:coupling_eq2}) naturally ensue and can be studied. The nonlinear evolution of plasma waves coupled with the ion dynamics, which is the basis for LDI, has already been evidenced in previous laser laboratory experiments \cite{Riconda2011,Depierreux00a,depierreux02a,Michel10a}. Furthermore, laser-generated plasmas can cover a large range of electron densities and temperatures, allowing to get conditions very similar to space plasmas \cite{Robinson93a} in terms of the dimensionless quantities that govern the nonlinear evolution of LWs and the amplitude and angular pattern of the EM $2\omega_p$ radiation: in laser-based experiments $\kLW\lD \sim 0.01-0.3$, $\kEM/\kLW \sim  0.15-0.65$, $T_e/T_i > 1$. Here $\kLW$ and $\kEM=\sqrt{3}\omega_p/c$ are the wavenumbers of the LW and the EM wave, $\lD$ is the Debye length, $T_{e,i}$ the electron and ion temperatures. In the interplanetary space (near-Earth orbit) $\kLW\lD \sim 0.014-0.05$, $\kEM/\kLW \sim  0.17-0.5$, while electron beam-based experiments have covered a lower range of $\kEM/\kLW$ (0.07 to 0.19).

In this Letter, we present the first experimental evidence of EM emission at $2\omega_p$ from laser interaction with an underdense plasma. The primary Langmuir wave is excited by a laser via stimulated Raman scattering (SRS) \citep{Drake1974,Forslund1975,Kruer1988}, a 3-wave resonant process between the laser EM wave, the LW, and a scattered EM wave. Energy and momentum conservation results in the following equations on frequencies and wavevectors: $\omega_0 = \omega_p + \wSRS$ and $\vkO = \vkLW + \vkSRS$, where the indices $0$, LW and SRS denote the laser, Langmuir, and scattered waves, respectively. Since the growth rate of SRS is highest for backward scattering (of the EM wave), the primary LW has a wavevector $\vkLW$ with an orientation close to that of the laser ($\vkO$).

\begin{figure}
		\includegraphics[width=\columnwidth]{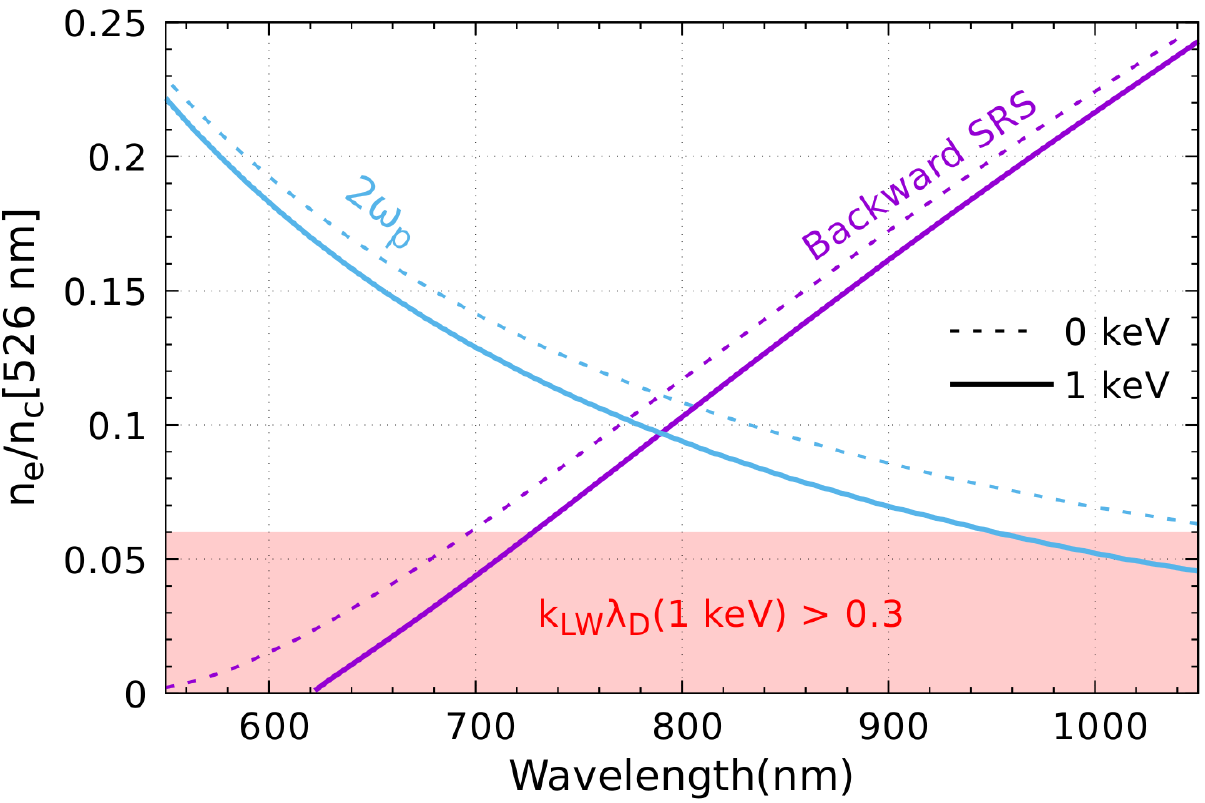}
		\caption{\label{SRS_2w_wavelength_vs_ne} Correspondence between the electron plasma density $n_e$ (expressed in $n_c$, for $\lambda_0$ = 526 nm) and the wavelengths of the backward Raman scattering and the $2\omega_p$ emission, for two electron plasma temperatures, 0 and 1 keV. The red shaded area corresponds to the density range where the backward Raman LW is strongly Landau damped (factor $\kLW\lD > 0.3$).}
	\end{figure}

The frequency of the Raman scattered light ($\omega_0 - \omega_p$) and  that of the EM harmonic plasma emission ($2\omega_p$) have opposite behaviors as a function of the plasma density, as depicted in Fig. \ref{SRS_2w_wavelength_vs_ne}, for $T_e$ = 1 keV (the same figure shows the weak dependence on $T_e$). In our experiment, we probe a plasma profile that is expanding in time, resulting in a decreasing density (decreasing $\omega_p$). We thus identify the two emissions from the opposite behavior of their frequencies \textit{vs.} time. The SRS instability grows only if $n_e/n_c < 0.25$ (where $n_c \sim 4\times 10^{21}$ cm$^{-3}$ is the critical density beyond which the plasma becomes opaque to the incident laser). It is inhibited by strong Landau damping when $\kLW\lD > 0.3$ \citep{LaFontaine1992}, corresponding to $n_e/n_c < 0.06$ for $T_e \sim$ 0.5-1 keV, so we do not expect emission in that range (shaded area). Note that the probed density range is limited by the spectral bandwidth of the diagnostics used: 680-880 nm, corresponding, for backward SRS, to $0.04 < n_e/n_c < 0.16$, and for the $2\omega_p$ emission to $0.07 < n_e/n_c < 0.14$.

	\begin{figure}
		\includegraphics[width=\columnwidth]{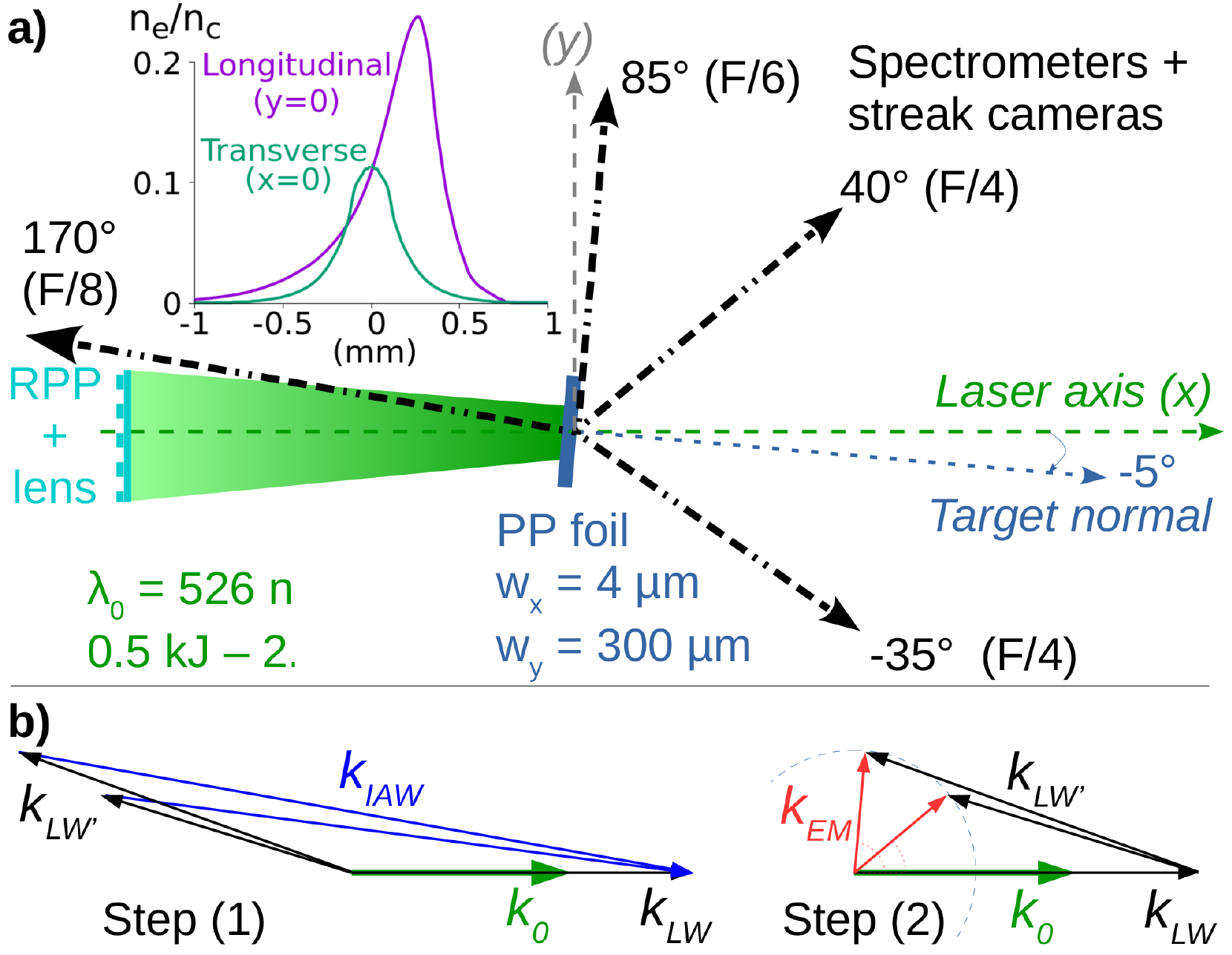}
		\caption{\label{Setup}a): Experimental setup, laser and target parameters, and typical longitudinal and transverse plasma profiles at the time of interest. b): wave-vector matching sketch showing one possible construction for the $2\omega_p$ emission at 85$^{\circ}$.}
	\end{figure}

Figure \ref{Setup}-a) outlines the geometry of the experiment, which was carried out at the LULI2000 laser facility (École Polytechnique, France). An expanding plasma is created by the irradiation of a 4 $\mu$m-thick, 300 $\mu$m-wide polypropylene (PP) foil with a 500 J, 2 ns-long laser pulse of wavelength $\lambda_0$ = 526 nm. The laser incidence angle is -5$^{\circ}$ from the target normal.  The 200 mm-diameter laser beam is focused using a random phase plate followed by a lens of focal length 1.6 m. The resulting focal spot has a speckle pattern forming a Gaussian spatial distribution with a full width at half maximum (FWHM) of $\sim$ 150 $\mu$m. The average intensity in the central region is $\sim 8\times 10^{14}$ W/cm$^2$, while the maximum speckle intensity is $\sim 5\times 10^{15}$ W/cm$^2$. The laser polarization is linear, with an angle of 45$^{\circ}$ with respect to the plane of incidence ($x$-$y$). The light scattered from the interaction region is collected in the same plane at 4 different angles from the laser axis (-35$^{\circ}$, 40$^{\circ}$, 85$^{\circ}$ and 170$^{\circ}$) by silver-coated spherical mirrors with respective F-numbers of 4, 4, 6 and 8. At each of these angles, the scattered light is spectrally and temporally resolved using an imaging spectrometer coupled to a streak camera. The covered spectral range is $650-900$ nm for the 170$^{\circ}$ and 85$^{\circ}$ diagnostics, and $700-850$ nm for the 40$^{\circ}$ and -35$^{\circ}$ diagnostics. For all diagnostics, the spectral and temporal resolutions are $\sim 10$ nm and $\sim 50$ ps. The interaction region is imaged on the entrance slit of each spectrometer. The probed volume is $\sim (100 \mu$m)$^3$ for the -35$^{\circ}$, 40$^{\circ}$ diagnostics, $\sim (175 \mu$m)$^3$ for the one at 85$^{\circ}$, and $\sim (400 \mu$m)$^3$ for the 170$^{\circ}$ diagnostic. The -35$^{\circ}$, 40$^{\circ}$ and 85$^{\circ}$ diagnostics are almost insensitive to the  polarization of the collected light. The spectral and energy responses of all diagnostics are absolutely calibrated, allowing for a sampling of the scattered light angular distribution at these four angles. Figure \ref{Setup}-b) shows wave-vector matching sketches showing steps (1) and (2) for the $2\omega_p$ emission at 85$^{\circ}$.

\begin{figure*}
		\includegraphics[width=\textwidth]{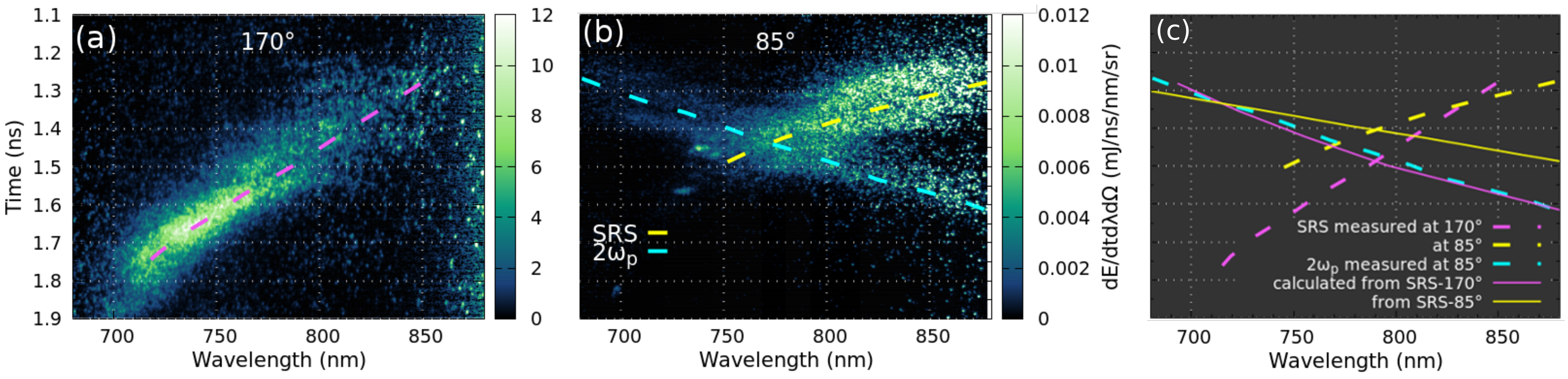}
		\caption{\label{spectra_vs_angle}Temporally and spectrally-resolved EM-emissions measured at (a) 170$^{\circ}$ and (b) 85$^{\circ}$ from the laser axis. The time t=0 corresponds to the beginning of the laser pulse. Central wavelengths are plotted as dashed lines. These lines are displayed again in panel (c), together with $2\omega_p$ wavelengths calculated from the 170$^{\circ}$ and 85$^{\circ}$ SRS signals (solid lines).}
	\end{figure*}

Typical temporally and spectrally-resolved measurements are presented in Fig. \ref{spectra_vs_angle}. Panel (a) illustrates the backward (170$^{\circ}$) spectrum, corresponding to SRS emission: its central wavelength decreases with time, in agreement with the expected density decrease from plasma expansion. No $2\omega_p$ emission is observed in the energy range of the detector, the signal is dominated by the SRS emission. Panel (b) shows the signal detected at 85$^{\circ}$ containing both the SRS and $2\omega_p$ emissions evolving in opposite ways. In agreement with Fig. \ref{SRS_2w_wavelength_vs_ne}, the two spectra cross at $\sim 780$ nm. In panel (c) we compare the experimental $2\omega_p$ detected at 85$^{\circ}$ with the harmonic frequency retrieved from both the 85$^{\circ}$ and 170$^{\circ}$ SRS signals. The harmonic frequency is retrieved as $2\omega_p = 2(\omega_0 - \omega_{\textrm{SRS}})$ at every time. From these three panels we observe that (i) the $2\omega_p$ wavelength follows more accurately the one retrieved from the 170$^{\circ}$-SRS. This is also true in terms of spectral intensity: (ii) both the 170$^{\circ}$-SRS and the $2\omega_p$ signals are weak at early times and strong at later times (see following Fig. \ref{spectra_vs_polar}), while the 85$^{\circ}$-SRS shows the opposite behavior. In addition, 
(iii) the $2\omega_p$ signal is present at late times when the 85$^{\circ}$-SRS has vanished. Finally, (iv) the SRS at 170$^{\circ}$ is 1000 times more intense than at 85$^{\circ}$. These four observations comfort the scenario of a $2\omega_p$ emission coming mainly from the LWs generated by backward SRS followed by LDI.

The origin of the $2\omega_p$ emission was further identified through its polarization. To characterize it on the signals collected at 85, 40 and -35$^{\circ}$, for some laser shots, a thin polarizer is placed in front of the spectrometer, retaining a polarization either parallel ($\parallel$) or perpendicular ($\perp$) to the plane of incidence. Results for both $\parallel$ and $\perp$ components are presented in Fig. \ref{spectra_vs_polar}. As expected, the SRS signal collected at 85$^{\circ}$ has a polarization mainly $\perp$: the projection of the laser polarization axis on the 85$^\circ$ diagnostic has a negligible $\parallel$ component. At smaller angles, the SRS signal is just above noise and no polarization analysis is possible. The $2\omega_p$ emission appears mainly $\parallel$-polarized. Assuming that the primary LW comes from the backward Raman instability, its wave-vector $\vkLW$ is parallel to the laser axis ($x$). In our set-up the measured EM$_{2\omega_p}$ wave has a wave-vector $\vkEM$ in the plane of incidence ($x$-$y$). Since the coalescence of the LWs must satisfy the momentum conservation relation $\vkLW+\vkLWp=\vkEM$, the wave-vector $\vkLWp$ of the secondary LW must lie in the same ($x$-$y$) plane. The longitudinal electric fields of both LWs are thus also in this plane, as must be that of the EM$_{2\omega_p}$ wave \citep{Gorbunov2004}, in agreement with our measurements. This aspect further confirms that the harmonic emission stems from the Raman-induced LW.

\begin{figure*}
	\includegraphics[width=\textwidth]{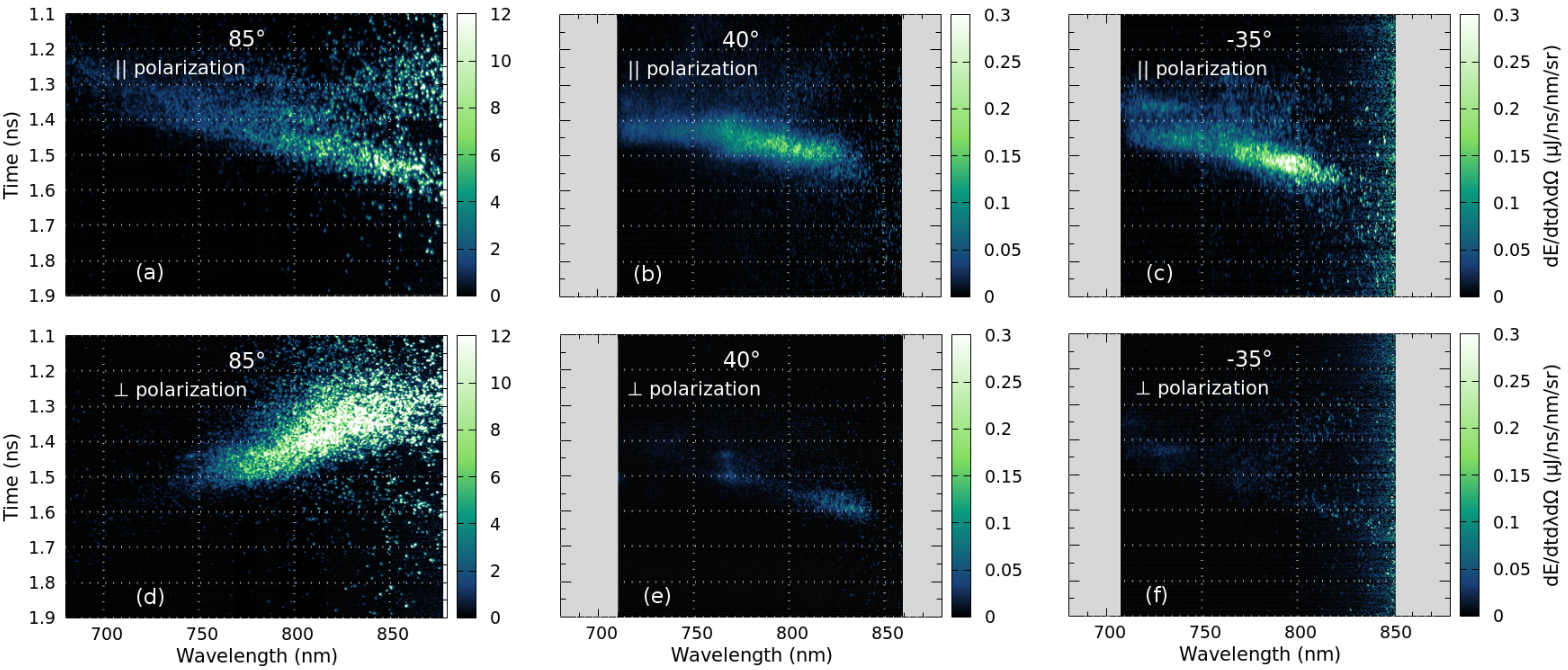}
	\caption{Temporally and spectrally resolved EM-emission measured at 85, 40, and -35$^{\circ}$. A polarizer placed in front of each diagnostic filters out either $\parallel$  or $\perp$  polarizations. Panels (a) to (c) correspond to $\parallel$-polarized emission, and panels (d) to (f) correspond to $\perp$-polarized emission. In (b) and (c) only the $2\omega_p$ signal is clearly observed, while in (d) only Raman is observed.}
	\label{spectra_vs_polar}
\end{figure*}

Let us now provide additional comments. Figure \ref{spectra_vs_angle}(a) shows that both the SRS and $2\omega_p$ signals vanish before 1.9 ns (before the end of the laser pulse at 2.5~ns, see also Fig. \ref{spectra_vs_polar}). This is caused by Landau damping which becomes dominant for $\kLW\lD \sim 0.3$, corresponding to $n_e/n_c \sim 0.06$ (see Fig. \ref{SRS_2w_wavelength_vs_ne}).
Furthermore, by spectrally integrating the $2\omega_p$ signal one can retrieve the instantaneous radiated power per solid angle. At 85$^{\circ}$ it reaches $\sim 0.3$ MW/sr. To estimate the possible attenuation of the $2\omega_p$ emission by the two-plasmon decay instability (TPD \citep{Kruer1988}, the inverse of LW coalescence) one can calculate an upper limit of the peak intensity of the $2\omega_p$ radiation at the source from the 85$^{\circ}$ signal: $3\times 10^7$ W/cm$^2$. For $T_e$ = 1 keV, $\lambda$ = 800 nm, and a density gradient of $\sim$ 100 $\mu$m, the intensity threshold for TPD is $\sim 6\times 10^{13}$ W/cm$^2$, thus ruling it out.

\begin{figure}
		\includegraphics[width=\columnwidth]{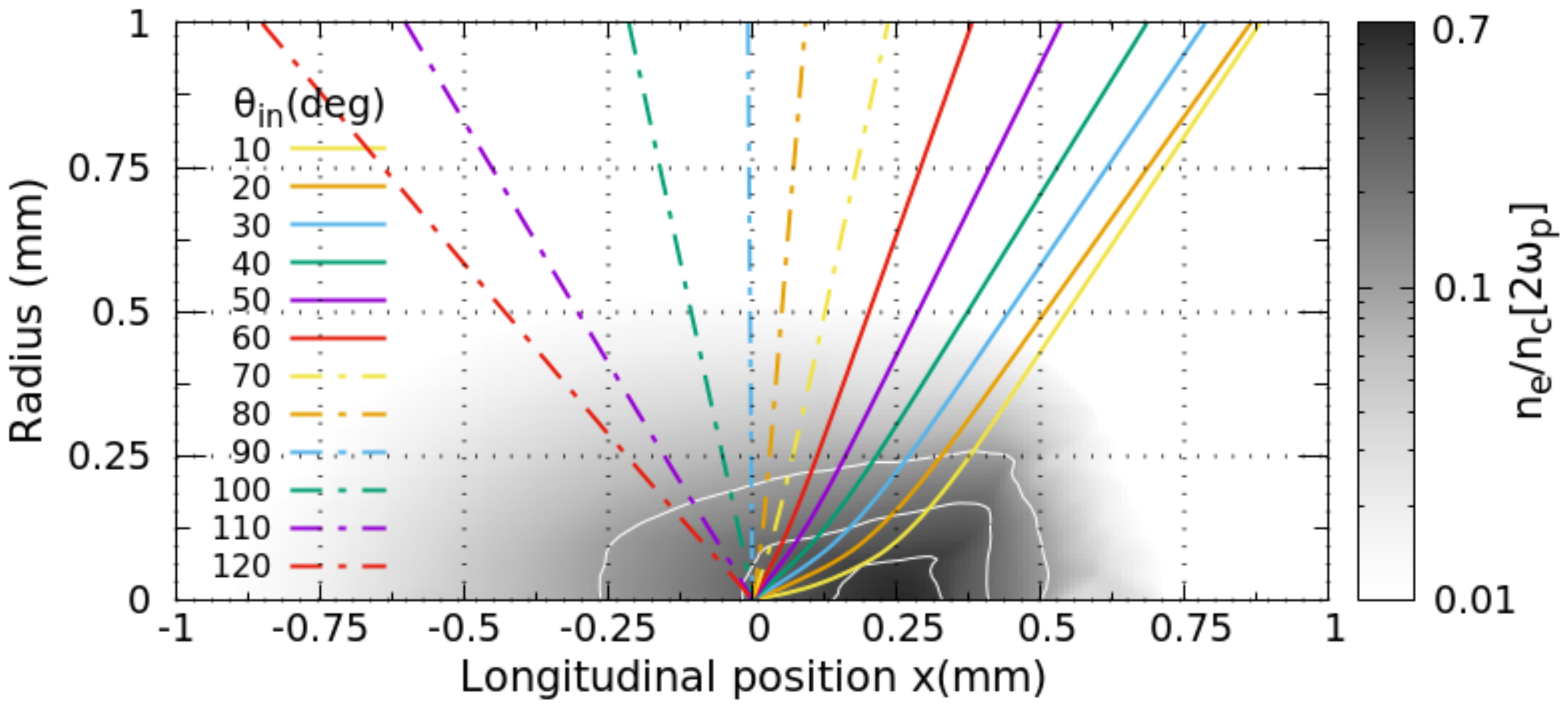}
		\caption{Electron density map from 2D cylindrical FLASH simulation (grey scale), and ray-tracing of $2\omega_p$ light emitted from the target center ($x=0,r=0$), showing the refraction calculated from the density map. White contour lines are at $n_e/n_c[2\omega_p]$ = 0.5, 0.25 and 0.1. The laser pulse is a square pulse of 2 ns duration and $5 \times 10^{14}$ W/cm$^2$ intensity. The focal spot size and the target correspond to the experimental ones. Time of the simulation is 1.3 ns from the pulse rising edge, so that the $2\omega_p$ emission from the target center (probed region) occurs at 760 nm, close to the measurements in Fig. \ref{spectra_vs_angle}-\ref{spectra_vs_polar}. At the target center $n_e \approx 4.3 \times 10^{20}$ cm$^{-3}$ and $T_e \approx$ 950 eV.}
		\label{simul_FLASH_ray_tracing}
	\end{figure}

The $2\omega_p$ emission is usually expected to have a quadrupole-like angular distribution, with minimum emission in the directions parallel and perpendicular to $\vkLW$ \citep{Henri19a}. The measured signals at 40$^{\circ}$ and -35$^{\circ}$ have similar intensities, as it would be expected from on-axis LWs generated by backward-SRS. Nevertheless, the $2\omega_p$ emission is $\sim$ 40 times less intense at 40$^{\circ}$ and -35$^{\circ}$ ($\sim$ 0.15 $\mu$J$/[$ns$\cdot$nm$\cdot$sr$]$) than at 85$^{\circ}$, in contradiction with the expected angular distribution. The 90$^{\circ}$-SRS might excite primary LWs at 45$^{\circ}$ from the laser axis leading to $2\omega_p$ emission peaked at 90$^{\circ}$. However, this was disproved by the previous scenario of $2\omega_p$ emission from backward-SRS-generated LWs, in particular the presence of $2\omega_p$ emission when 85$^{\circ}$-SRS is over.
In any case, to properly compare theoretical and experimental angular distributions, one should take into account the possible reflection, refraction, scattering and absorption of light in the surrounding plasma \citep{Steinberg1971,Itkina1992,Arzner1999,Thejappa2007}. To evaluate these effects, we first performed two-dimensional cylindrical hydrodynamic simulations (code FLASH, \citep{FLASH,FLASH2}) providing density and temperature profiles, then computed the propagation and absorption of $2\omega_p$ rays in the plasma gradients. A typical result is presented in Fig. \ref{simul_FLASH_ray_tracing}. The wave frequency ($2\omega_p$) is calculated by averaging the plasma density in a 100 $\mu$m x 100 $\mu$m cylinder centered at ($x$=0, $r$=0), equal to the experimentally probed volume. Rays starting with small initial angles $\theta_{in}$ are strongly refracted, which prevents rays from escaping with an angle below 40-50$^{\circ}$. In addition, their collisional absorption results in zero transmission for $\theta_{in}=0^\circ$ increasing up to 85\% for $\theta_{in}>60^\circ$. These results remain true for the whole duration of the $2\omega_p$ emission. 

The $2\omega_p$ energy collected by the $85^{\circ}$ diagnostic is of the order of 1.5 $\mu$J, in a solid angle of 0.022 sr. Even if one expects a quadrupolar emission, an estimate of the efficiency of the $2\omega_p$ conversion can be obtained assuming an isotropic emission. This leads to a total energy of $\sim$ 850 $\mu$J, thus a conversion efficiency of $\sim 1.7\times 10^{-6}$ from the laser. To estimate the conversion from LWs to the $2\omega_p$ emission, we compute the energy density contained in a LW of amplitude $\frac{\delta n_e}{n_e}$ as $W_{\rm{LW}} \approx \frac{1}{2} m_e c^2 n_e (\frac{\delta n_e}{n_e})^2$. At $n_e/n_c$ = 0.1 ($4\times 10^{20}$ cm$^{-3}$), assuming $\frac{\delta n_e}{n_e} \sim 10^{-2}$ in the probed volume leads to $W_{\rm{LW}} \sim$ 10 mJ. From the estimate above, the energy efficiency of the coalescence process is of the order of 10 $\%$.

In conclusion, the EM emission at twice the plasma frequency has been studied for the first time using laser interaction in underdense plasmas. Its spectral evolution, polarization, as well as intensity in three directions (-35$^\circ$, 40$^\circ$ and 85$^\circ$) have been characterized. They support the scenario of an EM wave generated by the coalescence of two LWs, with the primary wave excited by the backward Raman instability. The use of a plasma generated by laser irradiation of a solid target leads to large density gradients that deflect and absorb the EM radiation, strongly modifying the angular distribution. Symmetrical, low-gradient plasma produced from a laser-ionized supersonic gas jet would allow to reduce propagation effects. Even though the plasma parameters for the LW ($\kLW\lD$ and $k_{\rm{EM}}/k_{\rm{LW}}$) closely mimicked those found in the solar wind at near Earth orbit, the plasma was not magnetized. The EM wave is thus linearly polarized. Space-relevant magnetization ($\Omega_{ce}/\omega_p \sim 0.01$) could be achieved using fields of a few tens of Teslas, available from pulsed coils \citep{Albertazzi2013} or capacitive coils irradiated by a high-intensity pulsed laser \citep{Santos2015}.

Finally, let us note that in the context of inertial confinement fusion, EM-$2\omega_p$ emission could be a simple-to-implement diagnostic to retrieve information on SRS generated LWs and LDI. In addition, the efficient production of EM-$2\omega_p$ emission could provide an interesting path toward high intensity THz sources.

\begin{acknowledgments}
The authors would like to acknowledge the invaluable support of the LULI2000 staff. Financial support from CNES and Grant No. ANR-11-IDEX-0004-02 Plas@Par is acknowledged.
\end{acknowledgments}


\end{document}